\documentclass[12pt, reqno]{amsart}

\usepackage{latexsym, amsfonts, amsmath, amsthm, amscd} 
\newcommand{\ot}{\otimes}

\newcommand{\pair}[1]{\langle#1\rangle}

\newcommand{\boxt}{\Tox\kern -6.3pt\raise .55pt
    \hbox{$\scriptstyle{\times}$}}
 
\newcommand{\PP}{{\mathbb P}}

\newcommand{\ltimes}{\vbox to 5.4pt{\leaders\vrule\vfil}\kern
  -5pt\times}

\newtheorem{thm}{Theorem}[section]
\newtheorem{lem}[thm]{Lemma}

\newtheorem{prop}[thm]{Proposition}

\theoremstyle{definition}
\newtheorem{definition}[thm]{Definition}

\theoremstyle{remark}

\numberwithin{equation}{section}

\newcommand{\fg}{{\mathfrak g}} 
          
\newcommand{\fh}{{\mathfrak h}}
 
\newcommand{\fk}{{\mathfrak k}} 
    
\newcommand{\fl}{{\mathfrak l}}

\newcommand{\C}{{\mathbb C}}

\newcommand{\R}{{\mathbb R}}

\newcommand{\ga}{\gamma}

\newcommand{\eps}{\epsilon}

\newcommand{\la}{\lambda}

\newcommand{\CP}{\mathbb{CP}}

\begin{document}   

\title 
{Algebraic measures of entanglement} 
\author{Jean-Luc Brylinski}
\address{Department of Mathematics,
        Penn State University, University Park 16802}
\email{jlb@math.psu.edu}
\urladdr{www.math.psu.edu/jlb}
\thanks{I thank Joseph Bernstein and Ranee Brylinski for very useful discussions.}
\thanks{Research supported in part  by NSF  Grant No. 
DMS-9803593}
 

\begin{abstract} \hskip 4pc We study the rank of a general tensor $u$   in a tensor product
space\hfill\break
 $H_1\ot\cdots\ot H_k$. The rank of $u$ is the minimal number $p$ of pure states $v_1,\cdots,v_p$
such that
$u$ is a linear combination of the $v_j$'s. This rank is an algebraic measure of the degree of
entanglement of $u$. Motivated by quantum computation, we completely describe
the rank of an arbitrary tensor in $(\C^2)^{\ot 3}$ and give normal forms for tensor states up to
local unitary transformations. We also obtain partial results for
$(\C^2)^{\ot 4}$; in particular, we show that the maximal rank of a tensor in $(\C^2)^{\ot 4}$
is equal to $4$.

\end{abstract}
 
\maketitle 

\section{Rank of a tensor}
\label{sec:rank}
Let $H$ be  a complex Hilbert space; the hermitian scalar product will be denoted
by $\pair{u|v}$. It is complex linear in $v$ and antilinear in $u$.
 A state in a complex  $H$ is an element of the projective space $\PP(H)$.
The points of  $\PP(H)$ can be viewed alternatively as complex lines in $H$, or
as elements of the unit sphere $S(H)$ up to the scaling action of complex numbers
$e^{i\alpha}$. We will use the mathematical notation for states ($u$, $v$, etc...)
as opposed to kets. The state $u$ gives the
 projection operator $P_{u}:H\to H$  where
$P_{u}(v)= \pair{u|v}$. $P_u$ is an idempotent hermitian operator of rank $1$;
in this way we realize $\PP(H)$ as the orbit of the unitary group comprised of such operators.

In quantum mechanics, the combination of several quantum systems corresponds
to the Hilbert space tensor product
$E=H_1\ot \cdots\ot H_k$ of the relevant Hilbert spaces. A state $u$ in $E$ is called pure if it is a
tensor product
$\phi_1\ot\phi_2\cdots\phi_k$
 of states; otherwise it is called \emph{entangled}. It is easy to characterize
pure states in terms of homogeneous quadratic equations for the components of
the tensor $u$. If we pick orthonormal bases of each $H_j$ and write
$u_{a_1\cdots a_k}$ for the components of $u$, we have

\begin{prop} \label{prop:purity} The state $u$ in $E=H_1\ot \cdots\ot H_k$ is pure iff 
the following ``exchange property'' is verified: for any $k$-tuples
$(a_1,\cdots,a_k)$, $(b_1,\cdots,b_k)$, $(c_1,\cdots,c_k)$, $(d_1,\cdots,d_k)$
such that for each $j$, $(c_j,d_j)$ is a permutation of $(a_j,b_j)$ we have
\begin{equation}
u_{a_1,\cdots  a_k}u_{b_1,\cdots b_k}
=u_{c_1,\cdots  ,c_k}u_{d_1,\cdots d_k} 
\end{equation}

\end{prop}

\begin{proof}
Clearly a pure tensor satisfies the exchange property. To prove the converse, 
we proceed by induction
over $k$. We pick  a basis $(e_0,\cdots,e_m)$ of
$H_1$ and write $u=\sum_i~e_i\ot v_i$ where $v_i\in H_2\ot\cdots\ot H_k$. If $v_i\neq 0$
for some $i$, the exchange property for the case $a_1=b_1=i$ implies that $v_i$
satisfies the exchange property, so is a pure tensor by the inductive hypothesis.
Next, if $v_i$ and $v_j$ are non-zero, we apply the exchange property to the case where $a_1=d_1=i$,
$b_1=c_1=j$, $a_l=c_l$ and $b_l=d_l$ for $l\geq 2$, and conclude that the  tensors
$v_i$ and $v_j$ are proportional. It follows that $u$ is a pure tensor.
\end{proof}

Geometrically, the set of pure states is a closed complex algebraic subvariety
of $\PP(H)$, isomorphic to the product $\PP(H_1)\times\cdots\times \PP(H_k)$, which is
known as the Segre product. So its dimension is $d_1+\cdots+d_k-k$, where $d_j=\dim(H_j)$.
Accordingly, the pure tensors in $E=H_1\ot \cdots\ot H_k$ form a closed complex algebraic
subvariety of dimension $d_1+\cdots+d_k-k+1$.

Entangled states occur naturally in classical algorithms for matrix multiplication \cite{Str1}
\cite{Str2}.
They are very important in quantum mechanics; cf. e.g. the famous Einstein-Podolsky-Rosen work.
 Quantum computation lives in the tensor product Hilbert spaces $(\C^2)^{\ot n}$, and
states used in quantum coding  and quantum teleportation are typically quite entangled
(see e.g. \cite{C-R-S-S} \cite{Go} \cite{Ste}). So
it seems important to study how entangled  states can be. The following is a
classical notion (see \cite{B-C-S}).

\begin{definition} We say a state $u $ in $E=H_1\ot \cdots\ot H_k$
has rank $\leq p$ if we can write
\begin{equation}
u=\sum_{j=1}^p \lambda_j v_j
\end{equation}
where each $v_j$ is a pure state.

\end{definition}

A natural question is to find the highest rank of all states in $E$;
we can only answer this in very special cases. At least we can give a lower bound:

\begin{prop} \label{prop:lb}Let $H_j$ be vector spaces of dimension $d_j$. Then the highest
rank of states in $E=H_1\ot \cdots\ot H_k$ is at least equal
to the rational number
\begin{equation}
\frac{d_1d_2\cdots d_k}{d_1+d_2+\cdots +d_k-k+1}
\end{equation}

\end{prop}

For instance, take $k=3$, $d_1=3$, $d_2=4$, $d_3=5$; then the highest rank
is at least $3\times 4\times 5/10=6$.

In case $k=2$, it is easy to describe this degree of entanglement of any state
in classical terms:

\begin{prop} \label{prop:k=2}The degree of entanglement (rank) of a state $u$ in $E=H_1\ot 
H_2$ is the rank of the matrix $u_{ab}$.
\end{prop}

In particular, for $k=2$, the degree of entanglement  gives a nice stratification
of  projective space $\PP(E)$. Let $S_p$ denote the set of states of rank $\leq p$.
Then $S_p$ is a closed algebraic subvariety of $\PP(E)$, defined as the vanishing locus
of all order $p+1$ minors of the matrix $u_{ab}$. The singular locus of $S_p$ is
$S_{p-1}$.
The set
$S_p\setminus S_{p-1}$ of states of rank equal to $p$ is then a locally closed subvariety.

There is also a nice analytic characterization of pure states $\phi$ in $H_1\ot H_2$,
in terms of the projection operator $P_{\phi}$. The partial trace $\rho:=Tr_1(P_{\phi})$
is a positive hermitian operator on $H_2$ and we have:

\begin{prop}\label{prop:ana-purity}
We have $\rho^2\leq \rho$ with equality iff $\phi$ is pure.
\end{prop}

For the proof see Popescu-Rohrlich \cite{Po-Ro}. There is an interesting relation with the
algebraic characterization of pure states in Proposition \ref{prop:purity}, which we illustrate for
$\C^2\ot
\C^2$, using the basis
$(e_0,e_1)$ of $\C^2$. Here $\phi$ is given by a matrix $(\phi_{ab})$. The $(0,0)$-component
of $\rho-\rho^2$ is equal to $|\phi_{00}\phi_{11}-\phi_{01}\phi_{10}|^2$.
Thus the analytic equations characterizing pure states are quartic real polynomials
which are squares (in general, sums of squares) of absolute values of the quadratic complex
polynomial equations.

For $k>2$ the situation is more complicated: it is always true that $S_p\subset S_{p+1}$, but
we will see in the next section that the set
$S_p$ is not always  closed in $\PP(E)$. 

Note that for $k=3$ the rank of  a tensor is closely connected to the notion of rank  of a bilinear
map 
\cite{Str1}
\cite{Str2} \cite{B-C-S}.

\section{ Tensors in $(\C^2)^{\ot 3}$.}
\label{sec:3-qubits}

We study here $E=\C^{ 2}\ot \C^{ 2}\ot \C^{ 2}$. There is
a well-known polynomial function $D$ on $E$ which is invariant under $SL(2,\C)^3$:
this is the hyperdeterminant introduced by Cayley \cite{Cay} \cite{G-K-Z}.
$D$ is a  homogeneous degree $4$ polynomial function on $E$ which is $SL(2,\C)^3$-invariant.
We pick the standard basis $e_0,e_1$ of $\C^2$ and write the components of
$u\in (\C^2)^{\ot 3}$ as $u_{abc}$ for $a,b,c\in \{ 0,1\}$.
Then we have:

\begin{equation}
\begin{array}{ll}
D(u)=&u_{000}^2u_{111}^2+u_{001}^2u_{110}^2+u_{010}^2u_{101}^2+u_{011}^2u_{100}^2\\
&-2(u_{000}u_{001}u_{110}u_{111}+u_{000}u_{010}u_{101}u_{111}+u_{000}u_{011}u_{100}u_{111}\\
&+u_{001}u_{010}u_{101}u_{110}+u_{001}u_{011}u_{110}u_{100}+u_{010}u_{011}u_{101}u_{100})\\
&+4(u_{000}u_{011}u_{101}u_{110}+u_{001}u_{010}u_{100}u_{111})
\end{array}
\end{equation}

The geometric significance of $D$ is that $D(u)=0$ iff the hyperplane defined by $u$ is
tangent to the Segre product $S$ at some point $p$. This means that $\pair{u|v}=0$ 
for any tangent vector $v$ to $S$ at $p$.

 The review \cite{Cat} provides interesting comments
on the book \cite{G-K-Z}.

For a tensor $u$ in $E=(\C^2)^{\ot 3}$, there are three additional degrees of entanglement 
$\delta_1,\delta_2,\delta_3$ to consider: $\delta_1$ is the rank of $u$ viewed
as an element of $\C^2\ot \C^4$, when we group the second and third factors $\C^2$.
$\delta_2$ and $\delta_3$ are defined similarly.

We denote by $Y_j$ the closed algebraic subvariety of $\PP(E)$ comprised of states $u$ such that
$\delta_j=1$, i.e., $u$ belongs to $Y_1$ iff it is decomposable as a tensor in $\C^2\ot\C^4$.
Note each $Y_j$ has dimension $4$ and is contained in the hypersurface $Z$ of equation
$D=0$.

The following result is proved (at least implicitly) in \cite{G-K-Z}.
We include a proof since it uses methods which we will later use
for $(\C^2)^{\ot 4}$.

\begin{prop} \label{prop:3-qubits} Let $u$ be a state in $E= (\C^2)^{\ot 3}$. Then $u$ satisfies
exactly one the following possibilities:

(1) $u$ is a pure state.

(2) $u$ is not pure but belongs to $Y_j$ for a (unique) $j=1,2,3$.

(3) $u$ is entangled, and $D(u)\neq 0$; in that case $u$ has rank $2$, so it is the sum
of two pure tensors

(4)   $D(u)=0$, but $u$ belongs to none of the $Y_j$; then $u$ has rank $3$.

\end{prop}

\begin{proof}
It is useful to associate to $u$ a linear map $T:\C^2\to \C^2\ot \C^2$. 
If $T$ has rank $1$ then $u\in Y_1$ and we are in case (1). So we may assume $T$
has rank $2$. We will
consider $T(xe_0+ye_1)$ as a $2$ by $2$ matrix. 
Consider the homogeneous degree $2$ polynomial
$P(x,y)=det(T(xe_0+ye_1))$. There are $3$ cases to consider:

(I) there are exactly  two  points in $\CP^1$ where
$P$ vanishes. Let $(x_1,y_1),(x_2,y_2)$  be homogeneous coordinates for
these two points. Then we can make a change
a basis in the first
$\C^2$ so that these $2$ points are $(0,1)$ and $(1,0)$. Then both $T(e_0)$ and $T(e_1)$ have
rank
$\leq 1$; they must both be non-zero, otherwise all $T(xe_0+ye_1)$ would
have rank $\leq 1$. After a change  of basis in the second and third copies of
$\C^2$, we may assume $T(e_0)=e_0\ot e_0$ and $T(e_1)=e_i\ot e_j$ for suitable
$i,j$, not both equal to $0$. In this case, the tensor $u$ is equal to $e_0\ot e_0\ot
e_0+e_1\ot e_i\ot e_j$, so it has rank equal to $2$. By direct computation, we see that
$D(u)\neq 0$ if $i=j=1$ (case 4), or $u$ belongs to $Y_2$ (resp. $Y_3$) if $i=0$ (resp. $j=0$),
which belongs to case (2)..

(II) there is only one point $(x,y)$ of $\CP^1$ at which $P(x,y)$ vanishes. We may assume this
point is $(1,0)$. We can think of $T$ as giving a parameterization of a curve in $\CP^3$ which
is tangent to the quadric surface $Q$ consisting of rank $1$ matrices. We can change bases
in all three copies of $\C^2$ so that $T(e_0)=e_0\ot e_0$. As the tangent plane to $Q$ at
$e_0\ot e_0$ is spanned by  $e_0\ot e_1$ and $e_1\ot e_0$, we can change the basis vector $e_1$
in the first $\C^2$ so that
$T(xe_0+ye_1)=x
e_0\ot e_0+y(\la e_0\ot e_1+\mu e_1\ot e_0)$. Next, as $\la$ and $\mu$ must both be non-zero, we
can change bases in the other copies to arrange that $\la=\mu=1$.
Then our tensor $u$ is
$u=e_0\ot e_0\ot e_0+e_1\ot e_1\ot e_0+e_1\ot e_0\ot e_1$, and has rank exactly $3$.
Indeed it has the property that for any non-zero $v\in\C^2$, the tensor in $\C^2\ot\C^2$
obtained by contracting $u$ with $w$ has rank equal to $2$; thus $u$ can't be a sum
of two pure tensors.
We verify easily that $D(u)=0$. Or we can see geometrically  
that the corresponding point in
$\PP(E)$ belongs to the dual variety to the Segre product $S=S_1$, which means that the
hyperplane defined by $u$ is tangent to the Segre variety at some point. The
relevant point of $S$ is
$v=e_0\ot e_1\ot e_1$: notice that the tangent space to $S$ at $v$ is spanned by
tensors of the type
$\psi\ot e_1\ot e_1, e_0\ot\psi\ot e_1, e_0\ot e_1\ot\psi$ where $\psi\in\C^2$.
Since $u$ is orthogonal to all these tangent vectors, it is orthogonal to the tangent space
of $S$ at $v$. 

(III) the polynomial $P(x,y)$ vanishes identically; this means that the linear map
$T(xe_0+ye_1)$ always has rank $\leq 1$. This can happen in either of $2$ ways:

(a) there is a vector $\psi \in \C^2$ and a linear map $f:\C^2\to \C^2$ such that
$T(w)=\psi\ot f(w)$

(b) there is a vector $\psi \in \C^2$ and a linear map $f:\C^2\to \C^2$ such that
$T(w)=f(w)\ot\psi $

We need only consider case (a). Then we have $u=e_0\ot\psi\ot f(e_0)+e_1\ot\psi\ot f(e_1)$.
If $f(e_0)$ and $f(e_1)$ are linearly dependent, the tensor $u$ is pure and we are
in case (1). Otherwise, $u$ has rank $2$ and after changes of
bases in the second and third copies of $\C^2$ it takes the form $u=e_0\ot e_0\ot e_0
+e_1\ot e_0\ot e_1$. Then $u$ belongs to $Y_2$.

\end{proof}

This also leads to normal forms for tensor states in $(\C^2)^{\ot 3}$ up to the action
of $GL(2,\C)^3$; these normal forms are given in \cite{G-K-Z}. For quantum mechanics one needs
to consider the smaller symmetry group of unitary symmetries $U(2)^3$. This is the group of
local unitary symmetries; we say that two tensor states are \emph{locally equivalent} if they are
equivalent under
$U(2)^3$. One obtains  normal expressions up to local equivalence:

\begin{prop} \label{prop:normal_forms} A state in $(\C^2)^{\ot 3}$ is
locally equivalent to one of the following:

1) a pure state is locally equivalent to $e_0\ot e_0\ot e_0$.

2) a state in $Y_1$ which is not pure is  locally equivalent to

\begin{equation}
 e_0 \ot [\cos\theta ~ (e_0\ot e_0)+\sin\theta ~ (e_1\ot e_1) ]
\end{equation}

States in $Y_2$ or $Y_3$ are described similarly.

3) a state of rank 2 which is not in either of the $Y_j$ is locally equivalent to

\begin{equation}
\la ~ e_0\ot e_0\ot e_0 + z (\cos\theta_1e_0+\sin\theta_1e_1)\ot
(\cos\theta_2e_0+\sin\theta_2e_1)\ot (\cos\theta_3e_0+\sin\theta_3e_1)
\end{equation}
where $\la,\theta_j\in\R$, $z\in\C$  satisfy the relation
$\la^2+|z|^2+2\la \Re(z) \cos\theta_1 \cos\theta_2 \cos\theta_3=1$
(so that the tensor has norm $1$). We can assume $\theta_j\in (0,\frac{\pi}{2})$.

4) a state of rank 3 is locally equivalent to

\begin{equation}
\cos\theta_1 e_0\ot e_0\ot e_0 + \sin\theta_1
e_1\ot [\cos\theta_2 e_0\ot(\cos\theta_3 e_0+\sin\theta_3 e_1)+\sin\theta_2 e_1\ot e_0]
\end{equation}

\end{prop}

In each case there are only finitely many values of the parameters
corresponding a given tensor state.

\begin{proof} The four cases of the statement correspond to the four cases
of Proposition \ref{prop:3-qubits}. Case 1) is obvious. Case 2) follows as $u\in Y_1$
is locally equivalent to $e_0\ot v$ for some $v\in \C^2\ot\C^2$; by Schmidt's theorem
$v$ is locally equivalent to $\cos\alpha ~ (e_0\ot e_0)+\sin\alpha ~ (e_1\ot e_1)$.

In case 3), we have $u=v_1\ot v_2\ot v_3+w_1\ot w_2\ot w_3$ where $v_i$ and $w_i$ are
linearly independent for each $i$. There is no harm in assuming that the vectors
$v_1, v_2, w_1, w_2$ have norm $1$. By rescaling $u$ by a phase and applying
a local transformation  we can assume $v_1=v_2=e_0$ and
$v_3=\la e_0$ for $\la>0$. We can also arrange that $w_i=\cos\theta_ie_0+\sin\theta_ie_1$ for $i=1,2$
and $w_3=z[\cos\theta_3e_0+\sin\theta_3e_1]$ for some $z\in\C$. This gives the normal form;
note the reduction to $\theta_j\in (0,\frac{\pi}{2})$ is easy to achieve by changing the signs
of $e_0$ and $e_1$ in the $j$-th factor $\C^2$.

In case 4), the tensor $u$ has the form $u=v_1\ot v_2\ot v_3+w_1\ot(v_2\ot w_3+ w_2\ot v_3)$ 
where $v_i$ and $w_i$ are linearly independent for each $i$. There are two types of degrees
of freedom in the expression of $u$ in this form. First we have the transformation
$u=(v_1+\alpha w_1)\ot v_2\ot v_3+w_1\ot (v_2\ot[w_3-\alpha v_3]+w_2\ot v_3)$.
With its help  we can arrange that $w_1\perp v_1$. Secondly we have
$u=v_1\ot v_2\ot v_3+w_1\ot (v_2\ot[w_3+\beta v_3]+[w_2-\beta v_2]\ot v_3)$.
This is used to arrange that $w_2\perp v_2$. By rescaling the $v_i$, we may assume that $v_1$
and $v_2$ have norm $1$. After applying a local unitary transformation, we obtain
$v_1=v_2=e_0$ and $v_3=\la e_0$ for $\la\in\C^*$. Then we have $w_1=\alpha e_1$ and $w_2=\beta e_1$
for suitable $\alpha, \beta\in\C^*$. Write $w_3=\mu (\cos\theta e_0+\sin\theta e_1)$
for $\mu\in\C^*$. Thus we have
$u=\la e_0^{\ot 3}+\alpha e_1\ot (\mu e_0\ot[\cos\theta e_0+\sin\theta e_1]+\nu e_1\ot e_0)$
for some $\nu\in\C$.
Clearly a phase change for $u$ will make $\la$ real, so we can assume $\la\in\R$.
We can of course assume $\alpha=1$ by changing $\mu$ and $\nu$ appropriately.

In the rest of the proof we use the notation $e_i^{(j)}$ to denote the vector
$e_i$ in the $j$-th copy of $\C^2$. We will next do a simultaneous phase change 
\begin{equation}
e_0^{(1)}\mapsto e^{i\phi}e_0^{(1)}, e_0^{(3)}\mapsto e^{-i\phi}e_0^{(3)},
e_1^{(3)}\mapsto e^{-i\phi}e_1^{(3)}
\end{equation}
This operation does not change $\la$ but rescales $\mu$ as well as $\nu$; so we can assume $\mu$ is
real. Finally  a phase rescaling of $e_1^{(2)}$ will make $\nu$ real without changing
$\la$ or $\mu$. This way we easily get the normal form.

\end{proof}

It is interesting to discuss why $S_2$ is not closed in the case of $(\C^2)^{\ot 3}$. There is
an easy geometric description of the rank, which is well-known to algebraic geometers.
We start with the Segre product $S=S_1$, which is a closed algebraic subvariety of 
$\PP(E)$. For a $(p-1)$-plane $\Pi\subset  \PP(E)$, we say that $\Pi$ is a
$p$-secant plane if $\Pi$ is spanned by $p$ points $y_1,\cdots,y_p$
of $\Pi\cap S$. For instance, a line is $2$-secant if it is a secant line, a $2$-plane
is $3$-secant if it spanned by $3$ points of $\Pi\cap S$. Then we have clearly

\begin{lem} \label{lem:secant} A point of $\PP(E)$ has rank $\leq p$ iff it belongs to some
$(p-1)$-plane $\Pi\subset \PP(E)$ which is $p$-secant to the Segre product $S$.
In other words $S_p$ is the union of all $(p-1)$-planes $\Pi$ which are $p$-secant.

\end{lem}

The point then is that $S_2$ need not be closed, because the limit of a sequence of 
secant lines to $S$ need not be a secant line, but could be a tangent line. This
is similar to the fact that the border rank of a bilinear map
can be lower than its rank \cite{B-C-L-R} \cite{Str2}. The same
phenomenon could occur for higher $p$. From algebraic geometry we have the following general
fact. In this statement, we use the Zariski topology of $\PP(E)$ for which the closed
subsets are the subsets defined by homogeneous polynomial equations. The constructible sets
are then those obtained from the Zariski closed subsets by finite Boolean
operations (finite unions, finite intersections, and complementation). A closed subset $F$ is
called irreducible if whenever $F=G\cup H$ for $G,H$ closed, we have $G=F$ or $H=F$.

Another interesting phenomenon is that a real tensor in $(\R^2)^{\ot 3}$ may have different
rank from the same tensor viewed as an element of $(\C^2)^{\ot 3}$. An example is the tensor
$e_0\ot (e_0\ot e_0-e_1\ot e_1)+e_1(e_0\ot e_1+e_1\ot e_0)$, which corresponds to the product
law $\R^2\ot \R^2\to \R^2$ on $\R^2=\C$; this has rank $3$ as a real tensor and rank $2$
as a complex tensor.

\begin{prop}  The set $S_p$ is a Zariski constructible subset of $\PP(E)$.
The closure $\bar S_p$ is irreducible.

\end{prop}

\begin{proof} Let $T_p$ be the image of an algebraic mapping $\Phi:X\to \PP(E)$,
where $X\subset S^p\times \PP(E)$ is the locally closed algebraic subvariety
comprised of $p+1$-uples $(x_1,\cdots,x_{p+1})$ where $x_1,\cdots,x_p\in S$ are distinct,
$x_{p+1}\in \PP(E)$ and $(x_1,\cdots,x_{p+1})$ belong to some $(p-1)$-plane,
and $\Phi(x_1,\cdots,x_{p+1})=x_{p+1}$. It is easy to see that $X$ is irreducible;
thus 
 standard results in algebraic geometry say that $T_p$ is constructible and its closure
is irreducible.
We have easily $S_p=\cup_{q\leq p}~T_q$ so that $S_p$ is constructible. It is clear
that $\bar T_q\subseteq \bar T_p$ for $q\leq p$, so that $\bar S_p=\bar S_p$
is irreducible.

\end{proof}

The method of proof of Proposition \ref{prop:3-qubits} leads naturally to the following notion 

\begin{definition} Let $F$ be subspace of $E=H_1\ot \cdots\ot H_k$. The rank of $F$
is the smallest integer $p$ such that there exist $p$ pure tensors $u_1,\cdots,u_p$
such that $F$ is contained in the span of $u_1,\cdots,u_p$.
\end{definition}

We then have the following easy but useful result:

\begin{lem} \label{lem:ent_ss} \cite[Prop. 14.44]{B-C-S} Let $u\in E=H_1\ot H_2\ot\cdots\ot H_k$,
and let
$T$ be the corresponding linear map $T:H_1^*\to H_2\ot \cdots\ot H_k$. Then the rank of the tensor
$u$ is equal to the rank of the range of $T$ as a subspace of $H_2\ot \cdots\ot H_k$.

\end{lem}

\begin{proof} Let $p$ be the rank of $u$ and $q$ the rank of the range of $T$.
Thus $u$ is a linear combination of pure tensors $v_1,\cdots,v_p$. Write
$v_j=w_j\ot z_j$ where $w_1\in H_1$ and $z_j\in  H_2\ot \cdots\ot H_k$.
Then we have $T(l)=\sum_j ~ \pair{l|w_j}z_j$,  so that $T(l)$ is a linear
combination of the pure tensors $z_j$ and $q\leq p$. In the other direction, assume
that the range of $T$ is contained in the linear span of the pure tensors
$\beta_j,1\leq l\leq s$. Then there are linear forms $v_j$ on $H_1^*$ (so $v_j\in H_1$)
such that $T(l)=\sum_{j=1}^s~\pair{l|v_j}\beta_j$. This means that
$u=\sum_{j=1}^s~v_j\ot\beta_j$ and $r\leq s$.

\end{proof}

There is a classical example for the rank of a subspace of $M_2(\C)^{\ot 2}$.
We identify $M_2(\C)$ with its dual, so that $M_2(\C)^{\ot 2}$ identifies with the space of
bilinear functionals $(A,B)\mapsto f(A,B)$ of two matrices $A,B$ of size $2$.
The coefficients of the product $AB$ yield four such bilinear functionals, which span
a four-dimensional subspace $E$ of $M_2(\C)^{\ot 2}$. It is a well-known result
of Strassen \cite{Str1} \cite{Str2} that the rank of this subspace is equal to $7$ (instead
of the value $8$ one might naively expect). This is the basis for fast matrix
multiplication. From  Lemma \ref{lem:ent_ss} it
ensues that the corresponding tensor in $M_2(\C)\ot M_2(\C)\ot M_2(\C)=M_2(\C)^{\ot 3}$ has
rank equal to $7$. It would be nice to have a geometric interpretation
of this fact.

We also note an easy consequence of Lemma \ref{lem:ent_ss}

\begin{lem} \label{lem:ub} Let $(e_1,\cdots e_{d_1})$ be a basis of $H_1$, and consider
a tensor $u=\sum_j~ e_j\ot v_j\in H_1\ot H_2\ot\cdots\ot H_k$. Then
the rank of $u$ is at most the sum of the ranks of the $v_j$'s.
\end{lem}

\section{ Tensors in $(\C^2)^{\ot 4}$.}
\label{sec:4-qubits}
Our results for $E=(\C^2)^{\ot 4}$ are less complete than for $(\C^2)^{\ot 3}$. 
For $E=(\C^2)^{\ot 4}$,
an important invariant is the following: for any permutation $(i,j,k,l)$ of $(1,2,3,4)$,
a tensor $u\in (\C^2)^{\ot 4}$ yields a linear map 
  $\phi_{ijkl}:\C^2\ot\C^2\to \C^2\ot\C^2$. We consider the determinant
$\Delta(ijkl)=det(\phi_{ijkl})$. We have the following symmetries:
$\Delta(ijkl)=-\Delta(jikl)=-\Delta(ijlk)=\Delta(klij)$, so up to sign
 we have essentially $3$ determinants.
Now in fact we have

\begin{lem} \label{lem:deltas}
\begin{equation}
\Delta(1234)-\Delta(1324)+\Delta(1423)=0.
\end{equation}

\end{lem}

\begin{prop} \label{lem:S3} Let $E=(C^2)^4$. Then the closure of $S_3$ in $\PP(E)$ is the
dimension
$13$ algebraic variety defined by the equations $\Delta(ijkl)=0$
\end{prop}
\begin{proof} The Lie group $G=GL(2,\C)$ acts naturally on $E$ and preserves each $S_p$.
Let $T$ be the subspace spanned by $e_0^{\ot4}$,  $e_1^{\ot 4}$ and $(e_0+e_1)^{\ot 4}$.

Clearly the closure of $S_3$ is the closure of the $G$-saturation $G\cdot T$.
We can compute its dimension as follows. We consider the infinitesimal
equation of the Lie algebra $\fg=\fg\fl(4,\C)$ on $E$. For $v\in T$, we denote by
$\fh_v$ the space of $\ga\in\fg$ such that $\ga\cdot v\in T$. Then we have

\begin{equation}
\dim(G\cdot T)=\dim(G)+\dim(S)-min_{v\in S} \dim(\fh_v)-1=18-min_{v\in S} \dim(\fh_v)
\end{equation}

This follows as the right-hand side is the rank of the mapping $G\times T\to T\to\PP(E)$
at the point $(1,v)$.

Now for any $\delta,\eps\in \C^*$, the tensor $u=u_{\delta,\eps}=e_0^{\ot
4}+\delta e_1^{\ot 4} +\eps (e_0+e_1)^{\ot 4}$ belongs to $T$. Let $\fk_u$ be the space
comprised of the $\ga\in\fg$ such that $\ga\cdot u$ is a linear combination
of $e_0^{\ot 4}$ and $e_1^{\ot 4}$. Since $\fh_v$ is the direct sum of $\fk_v$
and of the line spanned by $(Id,0,0,0)$, we have $\dim(\fh_u)=\dim(\fk_u)+1$. So it suffices
to compute $\dim(\fk_u)$.

Now for $\ga=(\ga_j)\in \fg=\fg\fl(2,\C)^4$ with $\ga_j=\biggl(\begin{array}{ll}a_j&b_j\\
c_j&d_j
\end{array}\biggr)$  we compute:

\begin{equation}
\begin{array}{ll}
\ga\cdot u&=(a_1+a_2+a_3+a_4+\eps(a_1+a_2+a_3+a_4+b_1+b_2+b_3+b_4))~e_0^{\ot 4}\\
&+(c_4+\eps(a_1+b_1+a_2+b_2+a_3+b_3+c_4+d_4))~e_0^{\ot 3}\ot e_1+{\rm permutations}\\
&+(\delta d_1+\eps (a_1+b_1+c_2+d_2+c_3+d_3+c_4+d_4)) e_0\ot e_1^{\ot 3}+{\rm permutations}\\
&+\eps (a_1+b_1+c_2+d_2+c_3+d_3+c_4+d_4) e_0^{\ot 2}\ot e_1^{\ot 2}+{\rm permutations}
\end{array}
\end{equation}

So $\ga$ belongs to $\fk_v$ iff the coefficients of $e_0^{\ot 3}\ot e_1$, $e_0^{\ot 2}\ot
e_1^{\ot 2}$, $e_0\ot e_1^{\ot 3}$, and the other tensors obtained from these by permutations,
all vanish. At first sight this is just a system of linear equations in $16$ unknowns, but one
can essentially separate them according the four groups of four variables $(a_j,b_j,c_j,d_j)$
by introducing the sums $\la=\sum_j a_j,\mu=\sum_j b_j,\nu=\sum_j c_j, \rho=\sum_j d_j$.
 One gets the equations:

(1) For each $j$, $\eps(a_j+b_j-d_j)-(1+\eps)c_j=\eps (\la+\mu)$

(2) For each $j$, $\eps(a_j-c_j-d_j)+(\delta+\eps)b_j=\eps (\nu+\rho)$

(3) for each permutation $(ijkl)$ of $(1234)$ we have
$a_i+b_i+a_j+b_j+c_k+d_k+c_l+d_l=0$.

(3) easily implies that $a_j+b_j-c_j-d_j$ is independent of $j$.

By summing each of the three types of equations over all choices
of indices (or of permutations for the third), we get consistency requirements
for $(\la,\mu,\nu,\rho)$; these are easily solved to yield:

\begin{equation}
\mu=\frac{2\eps}{\delta-2\eps}\la,
\nu=\frac{-2\delta\eps}{\delta-2\eps}\la,
\rho=\frac{\la\delta(2\eps-1)}{\delta-2\eps},
\end{equation}

Here $\la$ is a free parameter; once it is chosen we can solve for $(a_j,b_j,c_j,d_j)$
 and obtain $\ga_j=\omega_j Id+\phi_j\xi+\eta$, where
$\xi=\biggl(\begin{array}{ll} \delta+\eps+\delta\eps &-\eps\\ \delta\eps&0\end{array}
\biggr)$, $\eta=\biggl(\begin{array}{ll} \frac{\delta\la}{\delta-2\eps} &0\\
0&0\end{array}
\biggr)$ are matrices independent of $j$, and $\omega_j,\phi_j$ are some scalars. The fact that
$a_j+b_j-c_j-d_j$ is independent of
$j$ then implies that
$\phi_j$ is too; call this scalar $\phi$.  Then we need the $a_j$ to 
sum up to $\la$, etc... This gives the value $\frac{-\mu}{4\eps}$ for $\phi$ and the requirement
$\sum\omega_j=\rho$.

Counting the free parameters we obtain $\dim(\fk_u)=4$. It follows that $S_3$ has dimension
$13$. It is clearly contained in the codimension $2$ subvariety defined by the vanishing
of the $\Delta(ijkl)$; the latter variety is seen to be irreducible, thus it must equal
the closure of $S_3$.

\end{proof}

\begin{thm} \label{thm:max_doe} The highest rank of a tensor in $(\C^2)^{\ot 4}$ is equal to
$4$.
\end{thm}

\begin{proof} We associate to $u\in (\C^2)^{\ot 4}$ as before a linear map $T:\C^2\to
(\C^2)^{\ot 3}$ and compute the rank of its image. If $T$ has rank $1$ it is clear that $u$
has rank
$\leq 3$, so we can assume $T$ is injective. We can think of $\phi$ as parameterizing
a line $l$ in $\PP((\C^2)^{\ot 3})=\C\PP^7$. If this line is not contained in the hypersurface
$Z$, then $2$ of its points have rank $\leq 2$, and it follows using Lemma \ref{lem:ub}
that $u$ has rank $\leq 2+2=4$. Thus we need to focus on the case where $l$ is contained in $Z$.
First of all, there is the case where $l$ is contained in $Y_j$ for some $j$.
In that case it is easy to see that $u$ is of rank $\leq 4$. So we can assume that $l$
contains a point $v$ which belongs to none of the $Y_j$; so $v$ is $GL(2,\C)^3$-conjugate to
$e_0\ot e_0\ot e_0+e_1\ot e_0\ot e_1+e_1\ot e_1\ot e_0$. For this choice of vector $v$ we can
write down the equations on a tensor $w$ so that the line thru $v$ and $w$ is contained
in $Z$, i.e., $D(xv+yw)$ vanishes identically. It is natural to consider
$w$ as a vector modulo scaling in the quotient space $(\C^2)^{\ot 3}/\C v=\C^7$, i.e.,
as an element of projective space $\C\PP^6$. Look at
the equation giving the vanishing of the coefficient of $x^iy^{4-i}$, as $i$ ranges from $3$ to
$0$. The first equation is
$w_{011}=0$. The second is $2(w_{111}^2+w_{001}^2+w_{010}^2)-(w_{001}+w_{010}+w_{111})^2=0$;
this is a non-degenerate quadratic form in $3$ variables. The third equation involves the new
variables $w_{000}, w_{101},w_{110}$ and is linear as a function of these $3$ variables.
The fourth equation involves also the last variable $w_{100}$, and is linear in $w_{100}$. The
variety of $w$ such that $D(xv+yw)\equiv 0$  thus has a dense open set which is obtained by
successive fibrations with fibers irreducible algebraic varieties; thus it itself is irreducible
and its dimension is equal to $2$. Denote by $Z^0$ the big $SL(2,\C)^3$-orbit inside $Z$, which
is the complement of $Y_1\cup Y_2\cup Y_3$. Now
consider the algebraic variety $S$ comprised of pairs
$(p,L)$ where $p\in Z^0$ and $L$ is a line thru $p$ which lies entirely inside $Z$. 
This is a locally closed subvariety of the product of $Z^0$ with the Grassmann
manifold of lines in $\C\PP^7$. Then the
projection map $S\to Z^0$ is a fibration, because it is $SL(2,\C)^3$-equivariant and $Z^0$
is  a single orbit. The dimension of $S$ is therefore $6+2=8$.
What we are really after however is the variety $V$ of lines contained
in $Z$  and meeting $Z^0$. There is an obvious map $S\to V$ which is a smooth mapping with
one-dimensional fibers.  Therefore $V$ has dimension $8-1=7$. Now
we claim that any line contained in $Z$ and not contained in any
$Y_j$ must meet each $Y_j$. For this purpose consider some tensor in $Y_1$, say 
$v=e_0^3+e_0\ot e_1^2$, and consider again the set of $w$ such that $D(xv+yw)\equiv 0$.

One checks that this forms a subvariety of $\PP^6$ of dimension $3$.
It follows that the set of lines contained in $Z$ and meeting $Y_1$ in finitely many points 
has a finite ramified covering which maps to
$Y_1$ with three-dimensional fiber, therefore it has dimension $4+3=7$. Note that the lines
completely contained in $Y_1$ form a  variety of dimension $5$. It then follows that any line
contained in $Z$ must meet each $Y_j$.

Thus we can change the basis of the first $\C^2$ so that $T(e_0)\in Y_1$
and $T(e_1)\in Y_2$. Then  both these tensors have rank $\leq 2$, and by Lemma \ref{lem:ub} $u$
itself  has rank $\leq 2+2=4$.

\end{proof}

It is easy to see that $S_2$ has dimension $9$ and satisfies a number of algebraic
equations, namely the $2$ by $2$ minors of the linear maps $(\C^2)^{\ot 2}\to (\C^2)^{\ot 2}$
obtained from the tensor (there are essentially $3$ such linear maps). It is likely the case
that these equations precisely describe the closure of $S_2$.

\end{document}